# An Example of the Quasiperiodic Solution of the Restricted Three Body Problem at 2:1 Resonance

Rosaev A.E.

*FGUP NPC Nedra, Svobody 8/38, 150000 Yaroslavl, Russia, rosaev@nedra.ru*

The 2:1 mean motion resonance orbit was integrated at the restricted planar 3-body problem in absolute frame. Orbit of Jupiter was assumed circular. Initial Jupiter longitude was assumed zero. The Runge-Kutta method was used. The start of first series of integration was from conjunction point at zero inclination and fixed eccentricity *e=0.4* and different pericenter longitudes.

The orbit with encounter at apocenter shows fast clockwise rotation. The orbit with encounter at pericenter rotated counterclockwise. It means, that periodic orbit exist between two investigated ones.

It was found, that two orbits with e=0.4, initial perihelion longitude ~ $\pm 100^{o}$ has not apsidal line rotation; however it has significant semimajor axis variations. The orbital elements show a very regular behavior on time interval about 3000 years. Due to Laplace theorem, at low perturbation, semimajor axis has only small short periodic oscillations. It means, that motion may be exactly periodic.

The cases of another initial eccentricity are considered at range from circular orbit to intersecting orbit. The dependence pericenter longitude of quasi-periodic orbits on eccentricity was found. The orbits with e > 0.5 have catastrophic close encounters with Jupiter and may be periodic only at special value of eccentricity. The additional series of integrations, at small shift from exact mean motion commensurability, was done.

## THEORETICAL INTRODUCTIONS

The apsidal line rotation rate (for planar case) can be estimated by averaging by Gauss:

$$\frac{d\omega}{dt} = -n - 2\sqrt{\frac{p}{fm}}\frac{dR}{dp}$$

$$R_1 = fm\left[(1) + (2)\left(\frac{e}{2}\right)^2 + (4)\left(\frac{e}{2}\right)^4\right]$$

The coefficients of expansion in according with Leverrier work [1] are:

$$(1) = \frac{A^0}{2} = \frac{b_1^0}{2a_1} \qquad (2) = a_1 \frac{dA^0}{da_1} + \frac{a_1^2}{2} \frac{d^2 A^0}{da_1^2}$$

$$(4) = 3a_1 \frac{dA^0}{da_1} + \frac{9}{2} a_1^2 \frac{d^2 A^0}{da_1^2} + \frac{3}{2} a_1^3 \frac{d^3 A^0}{da_1^3} + \frac{1}{8} a_1^4 \frac{d^4 A^0}{da_1^4}$$

Here $b_1^0$ - Laplace coefficients, $a_1$ - perturbing body orbit semimajor axis, n – mean motion, $p = a(1-e^2)$, $fm$ – perturbing body mass parameter.

It is evident, that, for special conditions, $\Delta\omega$ is possible to be zero.

NUMERIC INTEGRATION SETTING

Let now consider resonance 2:1 restricted three body problem with standard Jupiter mass. We starts from orbit with eccentricity = 0.4 at absolute frame. In any cases, the process of numeric integration starts at conjunction epoch. The Runge-Kutta method was used. Orbit with encounter at apocenter (fig1.) shows fast clockwise rotation. Orbit with encounter at pericenter (fig.2) shows counterclockwise rotation. It means, that orbit with fixed apocenter longitude exist between two investigated ones.

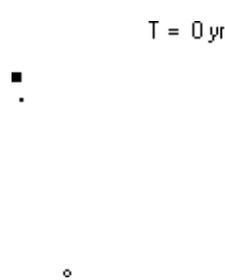
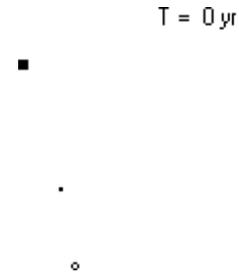

Fig.1. Initial configurations for integration. Conjunction/pericentre difference 180°.

Fig.2. Initial configurations for integration. Conjunction/pericentre difference 0°.

After that, a series of similar calculations was repeated at eccentricity e=0.3, e=0.2 and e=0.5.

# RESULTS

.
It is found, that orbit with e=0.4, perihelion longitude ~ 100° true anomaly - 100° has not apsidal line rotation; however it has significant semimajor axis variations.

By this way, the presence of quasi periodic solution close to this orbit is proved. Due to Laplace theorem, at low perturbation, semimajor axis has only small short periodic oscillations. It means, that motion may be exactly periodic.

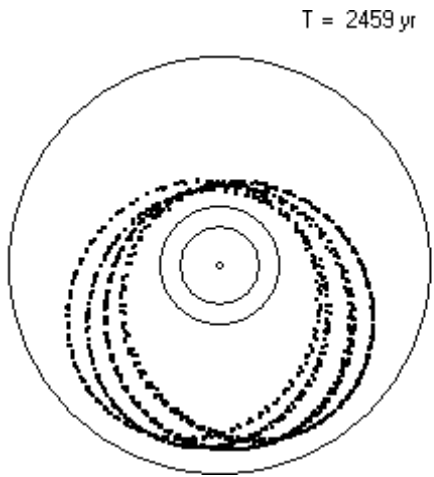
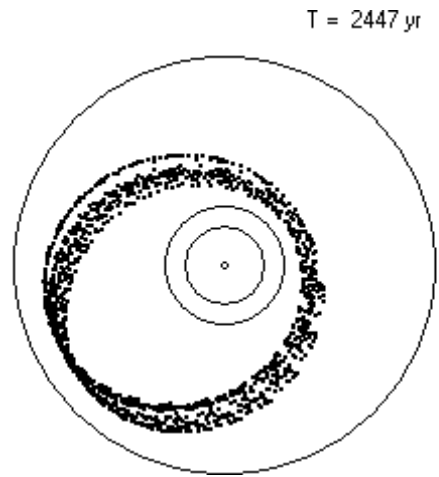

Fig.3. Longitude pericentre difference 50°. Counterclockwise orbit rotation

Fig. 4. Longitude pericentre difference 90°. Chaotic orbit variation

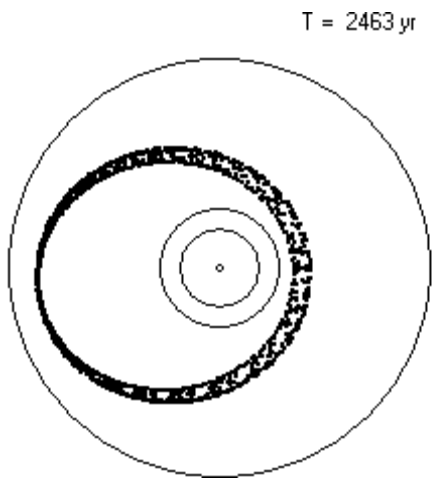
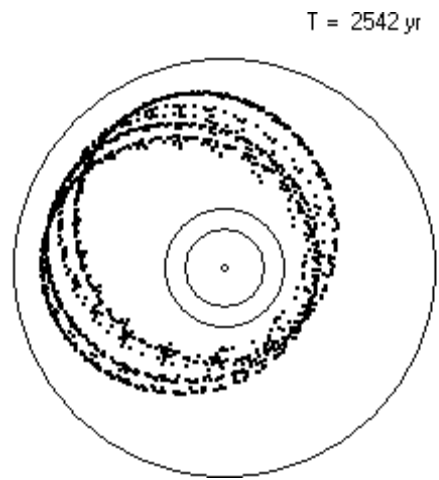

Fig.5. Longitude pericentre difference 100°. Quasi-periodic orbit

Fig.6. Longitude pericentre difference 110°. Clockwise orbit rotation

There are two orbits without apsidal motion as it is shown at fig. 7. As it is may be expected in agreement with fig. 9, the orbital elements variations are very regular at time interval about 3000 years.

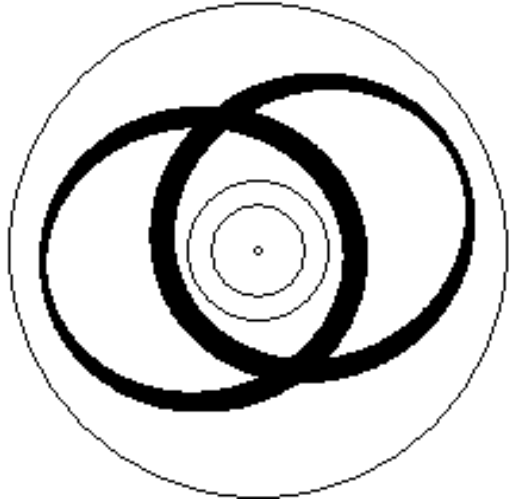

Table 1.
Quasi-periodic orbits at different eccentricity

| e   | Δλ  |
|-----|-----|
| 0.2 | 120 |
| 0.3 | 110 |
| 0.4 | 100 |
| 0.5 | 95  |

Fig.7. Two kinds of quasi periodic solutions

It gives a hope, that described periodic solution is stable. However, it is not possible to prove stability only by numerical experiments. It is possible to classify this solution as second kind Poincare solution. The shape of orbit in rotating frame, evidently is more complex [2]. At large eccentricity (e=0.5), the behavior of solution complicated by close encounter with Jupiter (fig. 9).

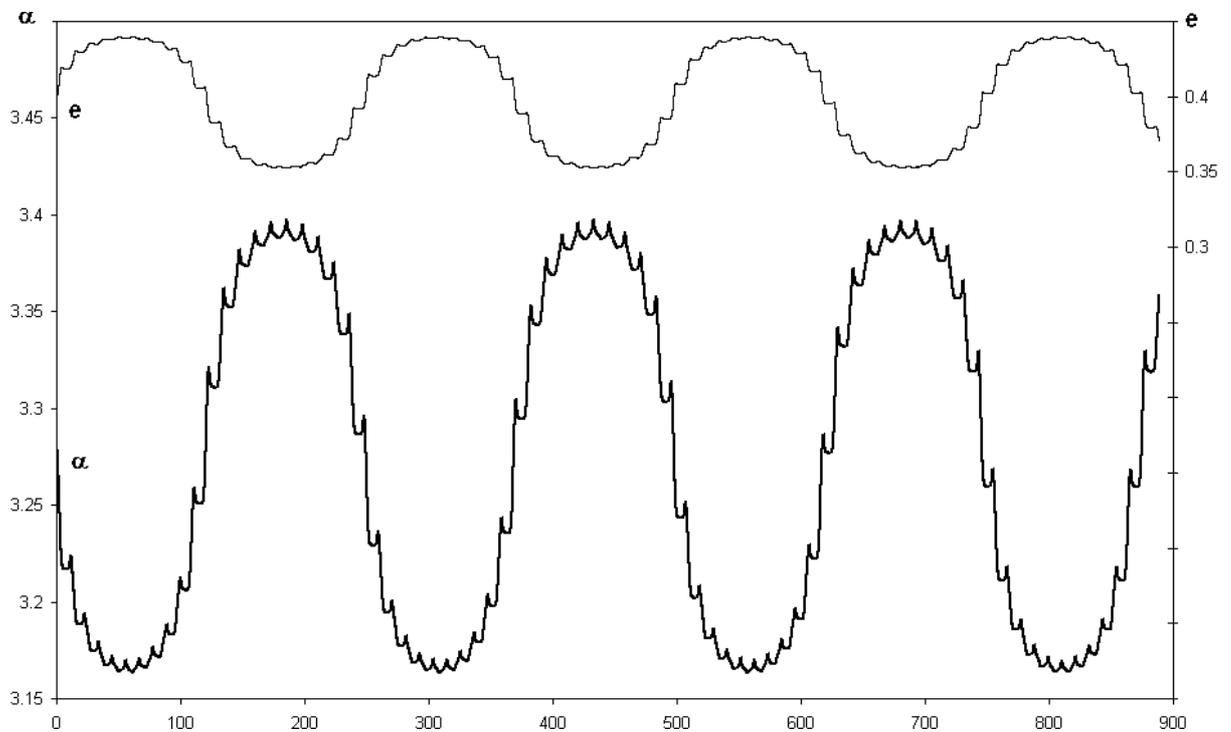

Fig.8. Evolution of semimajor axis and eccentricity for case $100°$ longitude pericentre difference

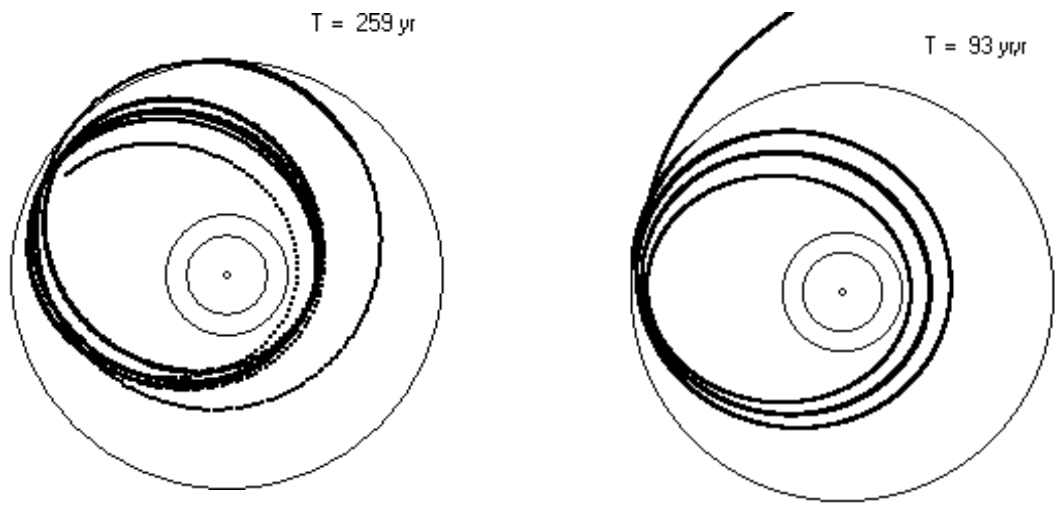

Fig.9. The instability of orbits with e=0.5 due to close enccounters

## CONCLUSIONS

The 2:1 mean motion resonance orbit was integrated at the restricted planar 3-body problem in absolute frame. Orbit of Jupiter was assumed circular. Initial Jupiter longitude was assumed zero. The Runge-Kutta method was used. The start of first series of integration was from conjunction point at zero inclination and fixed eccentricity $e=0.4$ and different pericenter longitudes.

The orbit with encounter at apocenter shows fast clockwise rotation. The orbit with encounter at pericenter rotated counterclockwise. It means, that periodic orbit exist between two investigated ones.

It was found, that two orbits with e=0.4, initial perihelion longitude $\sim \pm 100°$ has not apsidal line rotation; however it has significant semimajor axis variations. The orbital elements show a very regular behavior on time interval about 3000 years. Due to Laplace theorem, at low perturbation, semimajor axis has only small short periodic oscillations. It means, that motion may be exactly periodic.

The cases of another initial eccentricity are considered at range from circular orbit to intersecting orbit. The dependence pericenter longitude of quasi-periodic orbits on eccentricity was found. The orbits with e > 0.5 have catastrophic close encounters with Jupiter and may be periodic only at special value of eccentricity.